\documentclass[fleqn,twoside]{article}
\usepackage{espcrc2}

\usepackage{graphicx}
\usepackage[figuresright]{rotating}

\newcommand{\AmS}{{\protect\the\textfont2
  A\kern-.1667em\lower.5ex\hbox{M}\kern-.125emS}}

\title{Revealing treacherous points for successful light-front phenomenological 
applications}

\author{Chueng-Ryong Ji\address{Department of Physics, 
        North Carolina State University, \\ 
        Raleigh, NC 27695-8202, USA}%
        \thanks{Supported by Department of Energy under the contract 
DE-FG02-96ER40947.},
        Bernard L. G. Bakker\address{Department of Physics and Astronomy, 
        Vrije Universiteit, \\ 
        De Boelelaan 1081, NL-1081 HV Amsterdam, The Netherlands}%
        and
        Ho-Meoyng Choi\address{Department of Physics, Kyungpook National 
        University, \\
        Daegu 702-701, Korea}%
        \thanks{Supported by Korean Research Foundation under the 
contract KRF-2005-070-C00039.}}
       
\begin{document}

\begin{abstract}
Light-front dynamics (LFD) plays an important role in hadron
phenomenology as evidenced from recent development of generalized
parton distributions and other physical quantities involving hadrons.
For successful LFD applications to hadron phenomenology, however,
treacherous points such as zero-mode contributions should be taken into
account. For a concrete example of zero-mode contribution, we present
Standard Model analysis of vector anomaly in the CP-even form factors
of $W^\pm$ gauge bosons. The main distinguished features of LFD are
discussed in comparison with other Hamiltonian dynamics. We also
present a power counting method to correctly pin down which hadron form
factors receive the zero-mode contribution and which ones do not.
Indications from our analysis to hadron phenomenology are discussed.
\vspace{1pc}
\end{abstract}

\maketitle

\section{Introduction}

Light-front dynamics (LFD) provides a unified framework to analyze various
experimental measurements such as generalized parton distributions (GPDs)
and single spin asymmetry (SSA) at JLab and DESY (Hermes)\cite{GPV}, B-decays at 
SLAC (BaBar) and KEK (Belle)\cite{JC-BPhysics} as well as quark gluon plasma (QGP) 
productions at BNL (RHIC) and CERN (ALICE)\cite{Duke}, etc.. 
Due to the rational energy-momentum dispersion relation, the LFD has
distinguished features compared to other forms of Hamiltonian  
dynamics. In particular, the vacuum fluctuations are suppressed and the
kinematic generators are proliferated in LFD. Overall, these
distinguished features can be regarded as advantageous rather than as
disadvantageous in the hadron phenomenology.  However, in return, the
LFD implies treacherous points, an example of which one may realize from
the significance of zero-mode contributions even in the good (+) current
analyses\cite{BCJ-spin1}. Moreover, it has been shown that the common belief
of equivalence between the manifestly covariant calculation and the light-front (LF) 
calculation is not always realized\cite{BJ-endpoint} unless treacherous points are well taken
care of\cite{BDJM}. Thus, careful investigations of treacherous points and
judicious ways of handling those points should be precedent for LFD to
be distinctively useful compared to other forms of Hamiltonian
dynamics\cite{Dirac}.

In this presentation, we discuss a concrete example of a zero-mode contribution
found in the Standard Model analysis of the vector anomaly in CP-even form 
factors of $W^\pm$ gauge bosons. No model dependences are involved in this discussion
except that the analysis is made in the Standard Model. 
We then extend our discussion to hadron phenomenology.

In hadron phenomenology, observations of zero-mode contribution have been 
made by many authors\cite{BCJ-spin1,Jaus,BCJ-spin01,CJ-spin01,CJnew,Melo} for the 
electroweak form factors involving a spin-1 particle such as $\rho$ meson or deuteron. 
In particular, Jaus\cite{Jaus} proposed a covariant LF approach involving a lightlike
four-vector $\omega^\mu (\omega^2 = 0)$ as a variable and developed a way of finding
zero-mode contributions to remove spurious amplitudes proportional to $\omega^\mu$,
while our method of finding zero-mode contributions is a direct power counting
of the longitudinal momentum fraction for the off-diagonal elements in the Fock-state
expansion of the current matrix\cite{BCJ-spin1,BCJ-spin01,CJ-spin01,CJnew}.
However, Jaus\cite{Jaus} and we\cite{BCJ-spin1,BCJ-spin01,CJ-spin01,CJnew} do not agree on 
which form factors receive zero-mode contributions when the vector meson vertex 
$\Gamma^\mu$ is 
extended to more phenomenologically accessible ones. 
Utilizing manifestly covariant models for the vector meson
vertex $\Gamma^\mu$, we find that Jaus's method of finding zero-mode
contributions has a limitation for the choice of $\Gamma^\mu$\cite{CJ-spin01,CJnew}. 
In the absence of zero-mode contributions, the hadron form
factors can be obtained by just taking into account only the valence contributions
(or diagonal matrix elements in the LF Fock-state expansion). Thus, it is quite significant
in hadron phenomenology to correctly pin down which form factors receive the 
zero-mode contribution and which ones do not.

In the next Section, Section 2, we begin with a brief review on distinguished features
of LFD in comparison with other forms of Hamiltonian dynamics. In Section 3,
we present the zero-mode contribution in Standard Model analysis of vector anomaly
in CP-even form factors of $W^\pm$ gauge bosons. In Section 4, we discuss 
our power-counting method to pin down zero-mode contributions. Conclusions
follow in Section 5.
 
\section{Distinguished Features of LFD}

Among the three forms of Hamiltonian dynamics proposed by Dirac in 1949\cite{Dirac}, i.e.
instant ($x^0 =0$), front ($x^+ = x^0 + x^3 = 0$), point ($x_\mu x^\mu =
a^2 > 0, x^0 > 0$), the dimenson of stability group which leaves the
hypersurface of corresponding time invariant is the largest (i.e. seven out of the ten Poincare 
generators) in the front form which we call the LFD. Since the number of kinematic
operators is maximum among the possible forms of Hamiltonian dynamics, to a certain extent 
LFD is like sweeping dirt to a corner so that it leaves the rest of space 
clean\cite{sweepdirts,Ji-Bakker}. 
More reasons for our sweeping analogy may be due to the energy-momentum dispersion relation 
in LFD is 
rational; i.e.
\begin{equation}
p^- = \frac{{\vec p}_\perp^2 + m^2}{p^+},
\label{dr}
\end{equation}
where $p^- = p^0 - p^3$ is the LF energy (the conjugate variable
of the LF time $x^+ = x^0 + x^3$) and $p^+ = p^0 + p^3$ is the longitudinal
LF momentum orthogonal to $p^-$ for the particle with mass $m$ (or
$p^\mu p_\mu = m^2$). From Eq.(\ref{dr}), it is very clear that the signs 
of $p^+$ and $p^-$ are correlated. Thus, for all the particles and
antiparticles involved in the physical process, the LF longitudinal
momentum cannot be negative,i.e. $p^+ \geq 0$ in the LFD.
This leads to the simplicity of LF vacuum, i.e. the quantum fluctuation
of the vacuum is suppressed in LFD, except the zero-mode participation when all 
the particles and antiparticles in the physical process have zero LF
longitudinal momenta, i.e. $p^+ = 0$ for all individual constituents.

The simple vacuum except the zero-modes is a remarkable achievement in
LFD.  This is indeed like sweeping dirt to a
corner\cite{sweepdirts,Ji-Bakker} since the complexity of the vacuum is
condensed to the zero-mode contribution of the quantum states while the
rest of the vacuum is simple.  Among the various successful
applications of LFD utilizing the simplicity of vacuum, perhaps the
most well-known example may be the parton model in deep inelastic
lepton hadron scattering appreciated widely since 1970's.

However, the apparent simplicity of the LF vacuum yields a difficulty
in understanding the novel phenomena associated with nontrivial vacua
such as the spontaneous symmetry breaking, Higgs mechanism, chiral
symmetry breaking, axial anomaly, $\theta$-vacuum, etc..  To understand
these phenomena in LFD, we think that it is also very important to look
into dirts piled at the corner in our sweeping analogy.  First step in
this direction would be to look for the zero-mode contributions in the
vacuum phenomena. Realization of axial anomaly in LFD discussed in
Ref.\cite{Ji-Rey} was an attempt to this direction.  In the next
section (Section 3), we discuss an aspect of the vector anomaly in the
Standard Model and present a concrete example of a zero-mode contribution
to the vector anomaly.

\section{Vector Anomaly in Standard Model}

Anomalies betray the true quantal character of a quantized field
theory. Because they are invariably associated with divergent amplitudes,
their evaluation has proven to be complicated, at times even leading to
enigmatic results \cite{SV67}.
Nowadays there exists a vast literature on the subject and perhaps a
consensus has been reached \cite{Bert96}. By definition an anomaly is a
radiative correction that violates a symmetry of the classical Lagrangian
and usually involves counting infinities whether it is due to
ultraviolet infinities or an infinite number of degrees of freedom
\cite{Jac99}.
As this breaking of symmetry may bring quantized theory in agreement with
experiment, or, on the contrary spoil the renormalizability of the
theory, Jackiw \cite{Jac99} discerns with this distinction in mind two
types of infinities: good infinities and bad infinities.

In this presentation, we are concerned with the bad infinities which cause the
anomalies that ultimately spoil the predictive power of the theory.
In particular, we revisit the vector anomaly which led to the discussions
of the requirement of adding a contact term to the magnetic moment 
and the superconvergence relations \cite{DKMT}, {\it etc.},
in an effort to rescue the theory long time ago.
A brief historical remark on the anomaly associated with the
fermion-triangle loop was made in Ref.\cite{Ji-Bakker}.

The Lorentz-covariant and gauge-invariant CP-even electromagnetic $\gamma W^+ W^-$
vertex is defined~\cite{BGL,CN} by
\begin{eqnarray}
 \Gamma^\mu_{\alpha\beta} & = & i\,e \left\{ A[ (p+p')^\mu g_{\alpha\beta}
 +2(g^\mu_\alpha q_\beta - g^\mu_\beta q_\alpha)] \right.
\nonumber \\
 & &\left. + (\Delta \kappa)( g^\mu_\alpha q_\beta - g^\mu_\beta q_\alpha) \right.
\nonumber \\
 & &\left. + \frac{\Delta Q}{2M^2_W} (p+p')^\mu q_\alpha q_\beta\right\}, 
\label{eq.II.010}
\end{eqnarray}
where $p(p')$ is the initial(final) four-momentum of the $W$ gauge boson
and $q=p'-p$.  Here, $\Delta \kappa$ and $\Delta Q$ are the anomalous
magnetic and quadrupole moments, respectively. At tree level,
\begin{equation}
A = 1, \quad \Delta \kappa = 0, \quad \Delta Q = 0,
\end{equation}
for any $Q^2 = -q^2$ because of the point-like nature of $W^\pm$ gauge bosons.
Beyond the tree level, however,
\begin{eqnarray}
A &=& F_1(Q^2), \quad -\Delta \kappa = F_2(Q^2) + 2 F_1(Q^2), 
\nonumber \\
& &
\quad -\Delta Q = F_3(Q^2),
\end{eqnarray}
where $F_1, F_2$ and $F_3$ are 
the usual electromagnetic form factors 
for the spin-1 particles\cite{BJ02,BCJ-spin1}.
The physical form factors, charge ($G_C$), magnetic ($G_M$), and    
quadrupole ($G_Q$), are also related in a well-known way to the form
factors $F_1, F_2$ and $F_3$~\cite{BJ02,BCJ-spin1}.

Using the usual manifestly covariant technique of the dimensional   
regularization(DR) with $D=4-\epsilon$, which we denote as DR4, we
obtain:
\begin{eqnarray}
\noindent
&\left(F_2(q^2) + 2 F_1(q^2) \right)_{\rm DR4} = \frac{g^2 Q_f}{4 \pi^2}
\left\{\int^1_0 dx \int^{1-x}_0 dy \right.&
\nonumber \\
&
\left[(2-3x-3y)\right.\ln{C^2(m_1,m_1)}
+\left. \left. \frac{2 f^0_1 + f^0_2}{2 C^2(m_1,m_1)} \right]\right\},&
\nonumber \\
&\left(F_3(q^2)\right)_{\rm DR4} = -\frac{g^2 Q_f}{4 \pi^2}
\int^1_0 dx \int^{1-x}_0 dy\;&
\nonumber \\
&\frac{8xy(x+y-1)M^2_W}{C^2(m_1,m_1)},&
\label{eq.III.200}
\end{eqnarray}
where $C^2(m_a,m_b) =  xm^2_a + ym^2_b + (1-x-y)m^2_2 -
(x+y)(1-x-y)M^2_W -xy\;q^2$, $f^0_1 = -2[(x+y)(1-x-y)^2 M_W^2 + 
(2-x-y)xyq^2 + (x+y)m_1^2]$ and $f^0_2 = 2(x+y)[\{1-(x+y)^2\}M_W^2 +
xyq^2 +m_1^2]$.  Here, $m_1$ and $m_2$ are the masses of the charged
fermion struck by the photon and the spectator fermion, respectively. 
The charge factor $Q_f$ includes the color factor $N_c$ if the fermion
loop is due to the quark and $g^2 = G_F M_W/\sqrt{2} $ in the SM.  

The singular part involving $1/\epsilon$ of $\left( F_2 + 2F_1
\right)_{\rm DR4}$ vanishes as expected and  $F_3$ needs no
regularization. From Eq.~(\ref{eq.III.200}), we recover the well known 
$q^2 = 0$ results of the fermion-loop contribution to the physical  
quantities $\Delta \kappa$ and $\Delta Q$~\cite{BGL,CN}:

\begin{eqnarray}
(\Delta \kappa)_f & = &
 \frac{g^2 Q_f}{4 \pi^2} [I_4 + (F-2) I_3 + (2\Delta - {\cal E}) I_2],
\nonumber \\
(\Delta Q)_f & = &
 \frac{g^2 Q_f}{4 \pi^2} \;\frac{4}{3} [I_3 - I_4],
\label{eq.III.220}
\end{eqnarray}
where
\begin{equation}
I_m = \int^1_0 dt \frac{t^m}{t^2 -t F + {\cal E}},
\end{equation}
and $F= 1+{\cal E} -\Delta$, ${\cal E} = \frac{m_2^2}{M^2_W}$,
$\Delta=\frac{m_1^2}{M^2_W}$. 

In order to obtain the form factors other than by the standard DR
method, we may use the Pauli-Villars regularization method (PVR) or the
smearing method (SMR), which introduces fictitious particles with mass
$\Lambda$.  In the PVR, one replaces the amplitude $G^\mu(m_1,m_2)$ by
the amplitude $G^\mu_{\rm PV1} = G^\mu(m_1,m_2) - G^\mu(\Lambda_1,
m_2)$ or $G^\mu_{\rm PV2} = G^\mu(m_1,m_2) - G^\mu(m_1, \Lambda_2)$,
depending on whether one chooses to replace the propagator of the
charged fermion that is coupled to the photon or the spectator fermion
by the combination $(\not\!p- m)^{-1} - (\not\!p - \Lambda)^{-1}$. We
denote these regularization methods as PV1 and PV2, respectively.  The
smearing (SMR) procedure~\cite{BCJ01} consists in replacing the
photon-vertex $\gamma^\mu$ by the vertex $S_\Lambda (p') \gamma^\mu
S_\Lambda (p)$ using the smearing function $S_\Lambda (p) = \Lambda^2 /
(p^2 - \Lambda^2 + i\varepsilon)$.

The anomalous quadrupole moment $\Delta Q$ (or $F_3(Q^2)$) is found
completely independent from the regularization methods as it must be,
{\it i.e.}
\begin{equation}
\left( F_3 \right)_{\rm SMR} =
\left( F_3 \right)_{\rm PV1} =
\left( F_3 \right)_{\rm PV2} =
\left( F_3 \right)_{\rm DR4}.
\end{equation}
However, we find that the anomalous magnetic moment $\Delta \kappa$ (or
$F_2(Q^2) + 2F_1(Q^2)$) differs by some fermion-mass-independent
constants depending on the regularization methods:
\begin{eqnarray}
&\left( F_2 + 2F_1 \right)_{\rm SMR} -
\left( F_2 + 2F_1 \right)_{\rm DR4} = \frac{g^2 Q_f}{4\pi^2}\left(\frac{1}{6}
\right),&
\nonumber \\
&\left( F_2 + 2F_1 \right)_{\rm PV1} -
\left( F_2 + 2F_1 \right)_{\rm DR4} = \frac{g^2 Q_f}{4\pi^2}\left(\frac{2}{3}
\right),&
\nonumber \\
&\left( F_2 + 2F_1 \right)_{\rm PV2} -
\left( F_2 + 2F_1 \right)_{\rm DR4} = \frac{g^2 Q_f}{4\pi^2}\left(-\frac{1}{3}
\right).&
\label{cov-result}
\end{eqnarray}
These fermion-mass-independent differences are the vector anomalies
that we pointed out\cite{BJnew}.  Unless they are completely cancelled,
they would make a unique prediction of $\Delta \kappa$ impossible.
Within the SM, they are completely cancelled due to the zero-sum of the
charge factor ($\sum_f Q_f =0$) in each generation.

Since most calculations in the previous literature~\cite{DKMT,gww} were
limited essentially to use the manifestly covariant dimensional
regularization method, it has not yet been so clear how the appearance
of the vector anomaly changes in LF Hamiltonian formulation.  In LFD, the
Wick rotation to Euclidean space can never be used and the divergent
integrations are given only over the perpendicular components
$(k_x,k_y)$ of the integration variable after performing the LF-energy
($k^-$ pole) integration in Minkowski space.  The infinite parts in
$D=2-\epsilon$ can be expressed in the familiar form containing
$1/\epsilon$, which in the PVR and the SMR is replaced by a form
involving $\ln \Lambda$. We denote the LF regularization with
$D=2-\epsilon$ as DR2 for the distinction from the manifestly covariant
DR4 regularization. The details of the calculation were presented in
Ref.\cite{BJnew}, and in this presentation we note only the features
drastically different from the manifestly covariant case.

In LFD, we compute the form factors using the following relation
in the $q^+ = q^0 + q^3 =0$ frame,
\begin{eqnarray}
G^+_{++}&=& 2p^+(F_1 + \eta F_3),\;\nonumber \\
G^+_{+0}&=& p^+\sqrt{2\eta} (2F_1 + F_2 + 2\eta F_3),\nonumber\\
G^+_{+-}&=&-2p^+\eta F_3,\;\;\nonumber \\
G^+_{00}&=& 2p^+(F_1 - 2\eta F_2 - 2\eta^2 F_3),
\label{eq.II.100}
\end{eqnarray}
where $\eta = Q^2/4M^2_W$.
$G^+_{+-}$ depends on $F_3$ only and $G^+_{++}$ involves only
$F_1$ and $F_3$.  Therefore, the simplest procedure is to solve first
for $F_3$ from $G^+_{+-}$.  Next, $F_1$ is obtained from $G^+_{++}$ and
$F_3$. Finally, $F_2$ can be obtained from the other matrix elements. The
two relevant choices are to use  either $G^+_{+0}$ or $G^+_{00}$ and
consequently we may define
\begin{eqnarray}
&(F_2 + 2F_1)^{+0} = \frac{1}{p^+}\left[\frac{G^+_{+0}}{\sqrt{2\eta}} +
G^+_{+-} \right],&
\nonumber \\
&(F_2 + 2F_1)^{00} = \frac{1}{4p^+\eta}\left[(1+2\eta)G^+_{++} -
G^+_{00} \right.&
\nonumber \\
&\left. + (1+4\eta)G^+_{+-}\right].&
\end{eqnarray}

Splitting the covariant fermion propagator into the LF-propagating part
and the LF-instantaneous part, the divergences can show up both in the
valence amplitude containing only the LF-propagating fermions and in
the non-valence amplitude containing a LF-instantaneous fermion.  In
the $q^+=0$ frame, one might expect that the non-valence contribution
is absent since the integration range for the non-valence amplitude
diminishes to zero. However, this is not the case as we pointed out in 
Ref.~\cite{BCJ-spin1}. Calling the non-zero contribution from the non-valence
part in the $q^+ = 0$ frame the zero-mode, we find that only the
helicity zero-to-zero amplitude $G^+_{00}$ receives a zero-mode   
contribution and the non-vanishing zero-mode contribution to $G^+_{00}$
is given by
\begin{eqnarray}
&\left( G^+_{00} \right)_{\rm z.m.}
= \frac{g^2 Q_f p^+}{2\pi^3 M^2_W} \int^1_0 dx \int d^2 {\vec k}_\perp&
\nonumber \\
&\frac{{\vec k}^2_\perp + m^2_1 -x(1-x)Q^2}{{\vec k}^2_\perp + m^2_1 +x(1-x)Q^2}
\neq 0.&
\end{eqnarray}
The zero-mode contribution to $G^+_{00}$ is crucial because the
unwelcome divergences from the valence part due to the terms with a
power of the transverse momentum such as $(k_\perp^2)^2$ are precisely
cancelled by the same terms with the opposite sign from the zero-mode
contribution. Neither such high power $(k_\perp^2)^2$ term nor the
zero-mode contribution exist in $G^+_{+0}$. The essential results
directly related to the vector anomalies in DR2 are summarized as
follows:
\begin{eqnarray}
&(F_2 + 2F_1)^{+0}_{\rm DR2} - (F_2 + 2F_1)_{\rm DR4}
= \frac{g^2 Q_f}{4\pi^2} \left( \frac{1}{6} \right),&
\nonumber \\
&(F_2 + 2F_1)^{00}_{\rm DR2} - (F_2 + 2F_1)_{\rm DR4}
=& 
\nonumber \\
&-\frac{g^2 Q_f}{4\pi^2} \left(\frac{1}{2\eta}\right) \left(\frac{1}{3} +
\frac{2\eta}{9}\right).&
\label{LFanomaly}
\end{eqnarray}

The fact that $(F_2 + 2F_1)^{+0}_{\rm DR2}$ and $(F_2 + 2F_1)^{00}_{\rm
DR2}$ disagree indicates that the symptom of vector anomaly in DR2
appears as the violation of the rotation symmetry or the angular
momentum conservation ({\it i.e.} angular condition~\cite{CJ}).  This
appearance is drastically different from the case of the manifestly
covariant calculation. However, the anomaly-free condition ($\sum_f Q_f
= 0$) in the SM again removes the difference and restores the rotation
symmetry and the angular momentum conservation in LFD.

Besides DR2, we have also applied other regularization methods in LFD,
such as PV1, PV2 and SMR, which carry an explicit cutoff parameter
$\Lambda$.  Interestingly, in each of these regularization methods, we 
find that not only $(F_2 +2F_1)^{+0} = (F_2 + 2F_1)^{00}$ but also the
LF result completely agrees with the corresponding manifestly covariant
result: {\it e.g.}
\begin{equation}
(F_2 +2F_1)^{+0}_{\rm PV1} = (F_2 + 2F_1)^{00}_{\rm PV1} = (F_2 +2F_1)^{cov}_{\rm PV1},
\label{equivalence}
\end{equation}
where $(F_2 +2F_1)^{cov}_{\rm PV1}$ is the result shown in
Eq.~(\ref{cov-result}).  This proves that the rotation symmetry is
not violated in the regularization methods with an explicit cutoff
$\Lambda$ unlike the above DR2 case.  However, we note that the
zero-mode contribution in $(F_2 + 2F_1)^{00}$ is crucial to get
Eq.~(\ref{equivalence}). The details of our calculations including 
an interesting consequence in the PV2 case where the zero-mode is
artificially removed, were presented in Ref.~\cite{BJnew}. In
these LF regularization methods, we find that the vector anomalies
occur just the same way as in the manifestly covariant case (see
Eq.~(\ref{cov-result})).  Within the SM, they are again completely
cancelled due to the zero-sum of the charge factor ($\sum_f Q_f =0$) in
each generation.

\section{Power Counting Method}

For hadron phenomenology, Jaus~\cite{Jaus} and
we~\cite{BCJ-spin1,BCJ-spin01,CJ-spin01,CJnew} independently
investigated the spin-1 electroweak form factors in the past few
years.  As mentioned in the introduction, Jaus~\cite{Jaus} proposed a
covariant LF approach involving the lightlike four vector $\omega^\mu
(\omega^2 = 0)$ as a variable and developed a way of finding the
zero-mode contribution to remove the spurious amplitudes proportional
to $\omega^\mu$.  Our formulation, however, is intrinsically
distinguished from this $\omega$-dependent formulation since it
involves neither $\omega^\mu$ nor any unphysical form factors. Our
method of finding the zero-mode contribution is a direct power-counting
of the longitudinal momentum fraction in the $q^+\to 0$ limit for 
the off-diagonal elements in the 
Fock-state expansion of the current
matrix\cite{BCJ-spin1,BCJ-spin01,CJ-spin01,CJnew}.  Since the
longitudinal momentum fraction is one of the integration variables in
the LF matrix elements ({\it i.e.} helicity amplitudes), our
power-counting method is straightforward as far as we know the
behaviors of the longitudinal momentum fraction in the integrand.

For a rather simple (manifestly covariant) vertex $\Gamma^\mu
=\gamma^\mu$, both Jaus and we agree on the absence of zero-mode
contributions to the spin-1 electroweak form factors.
However, Jaus and we do not
agree when $\Gamma^\mu$ is extended to the more phenomenologically
accessible ones given by
\begin{equation}
\label{eq1}
\Gamma^\mu=\gamma^\mu
-\frac{(k+k')^\mu}{D},
\end{equation}
where $k$ and $k'$ are the relative four momenta for the two
constituent quarks.  Although Jaus's calculation and our calculation
used the same denominator $D$ in Eq.(\ref{eq1}), they led to 
different conclusions in the analysis of the zero-mode contribution.
Even if $D$ is chosen in such a way to get the manifestly covariant
$\Gamma^\mu$, the difference in the conclusions doesn't go away.

For the spin-1 elastic form factor calculations, Jaus's
conclusion\cite{Jaus} was that the matrix elements $<h'=0|J^+|h=1>$ and
$<h'=0|J^+|h=0>$ both get zero-mode contributions, so that one
cannot avoid the zero-mode contributions to the form factor $F_2(q^2)$
for the vector meson.  However, we found that only the matrix element
$<h'=0|J^+|h=0>$ gets a zero-mode contribution so that we can avoid
the zero-mode contribution to $F_2(q^2)$ if we do not use the matrix
element $<h'=0|J^+|h=0>$~\cite{CJ-spin01}.  

Similarly, for the weak transition form factors between the
pseudoscalar(P) and vector(V) mesons, Jaus\cite{Jaus} concluded that
the form factor $A_1(q^2)$[or $f(q^2)$] receives a zero-mode
contribution.  We again do not agree with his result but find that
$f(q^2)$ is free from the zero-mode contribution if the denominator $D$
in Eq.(\ref{eq1}) contains the term proportional to the LF energy
$(k^-)^{n}$ with the power $n>0$.  The phenomenologically accessible
LFQM satisfies this condition $n>0$.

As we have shown in detail in Ref.\cite{CJnew}, we can determine the
existence/nonexistence of the zero-mode contribution to $f(q^2)$ by
counting the factors of the longitudinal momentum fraction. For
example, if $D=D_{\rm cov}(k\cdot P)\equiv [2 k\cdot P + M_V(m_q+
m_{\bar q}) -i\epsilon]/M_V$, where $P$ is the four momentum of the vector
meson\cite{MF97}, then $D$ contains a term proportional to the LF
energy $(k^-)^{n}$ with the power $n = 1$.  This power-counting shows
that the form factor $f(q^2)$ should not receive a zero-mode
contribution in the $D_{\rm cov}(k\cdot P)$ case.  When the manifestly
covariant model for the vector meson vertex $\Gamma^\mu$ is available,
we have confirmed that the results found our way coincide with the ones
from the manifestly covariant calculation.

\begin{figure}[htb] 
\includegraphics[width=15pc,height=15pc]{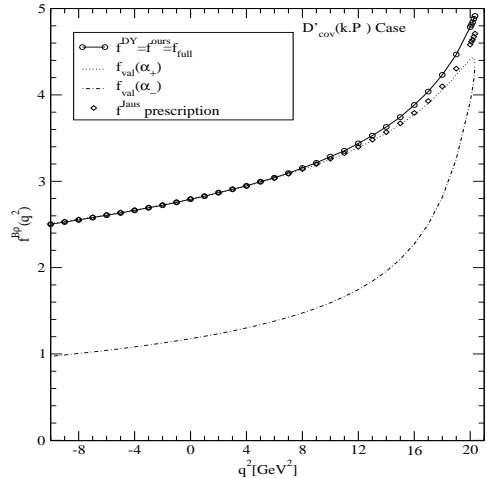} \caption{Weak
form factor $f(q^2)$ for $B \to \rho$ transition in the case of the
vector meson vertex with $D_{\rm cov}(k \cdot P)$.} \label{fig1}
\end{figure}

As we show in Fig.\ref{fig1}, our result $f^{\rm ours}$(circle)
obtained in the $q^+ = 0$ frame (or Drell-Yan frame so that $f^{\rm
ours}=f^{\rm DY}$) is in an exact agreement with the full result (solid
line) in the purely longitudinal $q^+ > 0$ frame\cite{CJnew}. This
agreement assures that our result is correct and there is no zero-mode
contribution to $f(q^2)$ for the vertex with $D_{\rm cov}(k \cdot P)$.
The nonvalence contributions in the purely longitudinal frame can be
found as the difference between the full result and $f_{\rm
val}(\alpha_+)$ (dotted line) or $f_{\rm val}(\alpha_-)$ (dotted-dashed
line) depending on the recoil direction of $B$-meson after it decays
into $\rho$ and leptons.

Comparing Jaus's result with ours, however, we find that $f^{\rm ours}$
and $f^{\rm Jaus}$ differ about 4\% at $q = q_{\rm max}$ although they
coincide at $q^2 =0$. This difference between Jaus's and ours shows
that Jaus's conclusion for the existence of zero-mode in $f(q^2)$
doesn't apply in this case $D = D_{\rm cov}(k \cdot P)$. Thus, Jaus's
method of finding zero-mode contributions has a limitation in the
choice of $\Gamma^\mu$.

\section{Conclusions}

In this presentation, we discussed a concrete example of a zero-mode
contribution in the Standard Model analysis of the vector anomaly computing
the CP-even form factors of the $W^\pm$ gauge bosons.  The vector anomaly in
the fermion-triangle-loop is real and shows a non-vanishing LF zero-mode
contribution to the helicity zero-to-zero amplitude $G^+_{00}$.  In LFD,
the helicity dependence of the vector anomaly exhibits a violation of Lorentz
symmetry.  
This may be contrasted with the manifestly covariant
Lagrange formulation where the same anomaly appears as a violation of
gauge symmetry. Our findings in this work may provide a bottom-up
fitness test not only to the LFD calculations but also to the theory
itself, whether it is an extension of the Standard Model or an
effective field theory of composite systems.

In the absence of zero-mode contributions, the hadron form factors can
be obtained rather straightforwardly by just taking into account only
the valence contributions. Moreover, the Lorentz  covariance of the
result can be assured in the absence of zero-mode contributions. For
successful hadron phenomenology, it is thus significant to pin down
which physical observables receive non-vanishing zero-mode
contribution.  Our power-counting-method discussed in this work seems
to provide a correct way of pinning down the existence/nonexistence of
the zero-mode contribution to physical observables.

\end{document}